# Cybersecurity for Autonomous Vehicles

Sai Varun Reddy Bhemavarapu

(Application Security Engineer), Texas, United States of America.

2**Introduction:**

The rise of autonomous vehicles has a significant transformation in the automotive industry. However, as these self-driving cars become more interconnected, their vulnerability to cybersecurity threats becomes a critical concern. While Safeguarding the security and integrity of autonomous systems is  important to prevent potential malicious activities that could jeopardize passenger safety, endanger other road users, and disrupt the entire transportation infrastructure.

The aim of this paper is to address the issues of cybersecurity within the autonomous vehicles by examining the challenges, risks which help in building our secure future. As these autonomous  vehicles rely on a intercommunication of sensors, artificial intelligence, communication systems, and external infrastructure which makes them vulnerable to various cyber threats. So It is crucial to address these vulnerabilities given the potential consequences they entail. A cybersecurity breach within the autonomous vehicle can have severe problems which can lead to loss of public trust and safety in autonomous technology. So, it is important to develop and implement best cybersecurity measures to ensure the viability and acceptance of self-driving cars on a large scale.

This paper analyses the key cybersecurity challenges that are faced by autonomous vehicles, including vulnerabilities in software and hardware components, risks associated with wireless communication, and external interfaces. It also explores existing cybersecurity frameworks and solutions tailored specifically for autonomous vehicles, such as secure software development practices, intrusion detection systems, Intrusion Prevention System, cryptographic protocols, and anomaly detection techniques. This paper also investigates the role of regulatory bodies, industry collaborations, and cybersecurity standards in helping a secure environment for autonomous vehicles. Establishing clear guidelines and best practices is essential to ensure consistent security measures across different manufacturers and jurisdictions, enabling a unified approach to addressing cybersecurity concerns.

By examining the current state of cybersecurity in autonomous vehicles and proposing effective countermeasures, this paper contributes to the advancement of the field. The goal is to promote the development and adoption of robust cybersecurity strategies that protect autonomous vehicles and passengers while inspiring public confidence in this transformative technology.

## What is an Autonomous vehicle?

**Definition**

Autonomous vehicles, are the vehicles that can drive themselves without any human control. While manufacturing the car, there are some special sensors like cameras, radar, and lidar which helps in gathering information about the environment around the vehicle have been installed. By Using this information from sensors, A special software inside the car creates a map of the environment and plans for a safe and efficient route. These Autonomous vehicles can also talk to other cars and infrastructure to share important information and work together accordingly. In simple terms, an autonomous vehicle is a car that can drive on its own using sensors and smart software (Parekh, et al., 2022).

**Background**

According to (M.Anderson, et al., 2016) In the year 1980s, Universities began to research on Autonomous Vehicles focusing on two types: one that relied on specific roadway infrastructure and another that do not require any road way infrastructure. United States Défense Advanced Research Projects Agency (DARPA) organized "grand challenges" to test the performance of Autonomous vehicles. In the year 2004, All Autonomous vehicles failed to complete 150-mile off-road course, but in the year 2005, five vehicles successfully finished. In the year 2007, six teams completed a 60-mile urban course following regular traffic rules and laws.

Apart from Autonomous vehicles, According to (Federal-aviation-administration, 2020) unmanned aircraft systems (UAS), which are now commonly known as drones, are being developed for various commercial applications. These include last-mile package delivery, transportation of medical supplies, and inspection of critical infrastructure. The integration of UAS into commercial ventures shows promising potential for enhancing efficiency and addressing specific industry needs.

**Levels of Automation**

The Society of Automotive Engineers (SAE) has established a scale with six levels to from level 0 to level 5. Level 0, where there is no automation and human driver is fully responsible for driving and control, to Level 5, where there is full automation. At Level 0, the human driver is entirely responsible for driving, even when using warning or intervention systems. In Level 1, a driver assistance system assists with steering or acceleration/deceleration, but the human driver remains responsible for all other driving aspects. Level 2 involves multiple driver assistance systems taking over steering and acceleration/deceleration, but the human driver is still responsible for all other driving tasks.

Moving to Level 3, the vehicle assumes control of all driving aspects, but the human driver must be ready to intervene if necessary. Level 4 signifies that the vehicle takes full control of driving, even without human intervention. Finally, at Level 5, the vehicle can handle all driving tasks under any roadway and environmental conditions that a human driver could manage.

The SAE automation levels provide a standardized way to describe the capabilities of autonomous vehicles. Researchers, manufacturers, and regulators utilize these levels to communicate and track the progress of autonomous vehicle development (Trasportation, 2021).

## What is Cyber security ?

**Definition**

"Cybersecurity is the practice of protecting critical systems and sensitive information from digital attacks. Cyber security is Also known as information technology (IT) security, cybersecurity measures are designed to combat threats against networked systems and applications, whether those threats originate from inside or outside of an organization" (IBM, n.d.).

**Importance of Cyber security**

Cybersecurity in today's digital ecosystems is very much important because it helps in protecting sensitive information, ensuring system integrity, and preventing unauthorized access to networks, computers, and data. As security breach could lead to huge financial losses, reputation damage to organisations, legal consequences, and loss of customer trust. By Implementing some of the effective practices, such as secure network infrastructure, encryption, access controls, and regular vulnerability assessments are very much important to defend against any cyber threats. Prioritizing employee training and keeping software and hardware updated are key in maintaining the confidentiality, availability, and reliability of digital assets.

## Cybersecurity in Autonomous Vehicles and its importance

Autonomous vehicles (AVs) are developing very quickly and also they rely on complex technologies, including sensors, software, and communication systems, to assess their surroundings and safely navigate and as a result, It is also important to focus on cybersecurity in this automotive industry to ensure the safety and security of passengers and other road users.

With all the dependence on software and communication networks these Autonomous Vehicles are more vulnerable to cyberattacks which can have catastrophic consequences, which include fatalities, injuries and other severe outcomes. Cyber threats targeting Autonomous vehicles can cause various risks, including data breaches, unauthorized access, and remote control manipulation. Moreover, securing communication channels and data exchange within and outside the vehicle is a critical aspect of AV cybersecurity.

To mitigate these risks, implementing good security measures is very much important. Some of the measures are strong encryption, authentication protocols, and intrusion detection systems are essential to protect data integrity and privacy. Continuous monitoring and proactive vulnerability assessments play a vital role in identifying and addressing potential weaknesses in AVs' cybersecurity defences. Putting good security measures in place is very crucial to mitigate these risks. A Strong encryption, authentication protocols, and intrusion detection systems are a few of the precautions that are very much necessary to protect the confidentiality and integrity of data. As detection and remediation of potential vulnerabilities in Autonomous Vehicles depend heavily on ongoing monitoring and regular vulnerability assessments.

A collaborative effort involving industry stakeholders, researchers, and regulatory bodies is needed to establish thorough cybersecurity standards and best practices. The safety and dependability of autonomous driving should be guaranteed by Autonomous

Vehicle manufacturers by complying to strict cybersecurity regulations, creating a safe environment for the widespread adoption of Autonomous Vehicles technology.

## Vulnerabilities And Threats In Autonomous Vehicles

**Potential Cyber Threats**

There are many potential cyber threats related to Autonomous vehicles. One of the potential cyber threats is the risk of unauthorized access to their control systems. This occurs when a malicious individuals tries to gain access into the vehicle's software or communication networks without permission of the user. The impact of such access could be very severe because it allows the attackers to manipulate or take control of the vehicle which can lead the lives of passengers and other road users at risk. For example, in the year 2015, researchers were able to remotely hack into a Jeep Cherokee, demonstrating the dangers of unauthorized access to AVs.

Another significant cyber threat is the possibility of data breaches in an autonomous vehicles. As these Autonomous vehicles collect and process large amounts of data which include sensor readings and personal information, a breach of this data could lead to privacy violations and even targeted attacks. Unauthorized access to the vehicle's navigation history or any type personal data stored within its systems could expose individuals to risks such as identity theft.

Denial-of-Service (DoS) attacks is also one of major potential threat in the autonomous vehicles. As, These attacks aim to disrupt normal vehicle operations by overwhelming communication networks or computational resources. By flooding the vehicle's systems with excessive traffic or requests, an attacker can make the vehicle unresponsive or cause it to behave uncommonly, endangering the safety of occupants and other road users. Imagine a situation where an AV's sensors are flooded with false signals which can result in incorrect driving decisions.

**Attacks based on Software Vulnerabilities**

As very well explained by (Jimenez & Amel, 2009), A software vulnerability is a flaw or defect in the software construction that can be exploited by an attacker in order to obtain some privileges in the system. It means the vulnerability offers a possible entry point to the system. There are many software vulnerbilities but some of the major attacks that can cause serious consequences are Denial of Service, Man in the middle attack, Remote code execution.

*Denial of Service*

This attack is also know as Dos attack. The aim of this type of attack is to disrupt or disable a system's normal operations by flooding the system with stream of requests which can exhaust system resources. The goal is to make the system unavailable to its intended users temporarily or indefinitely. Now when it comes to Autonomous vehicles then Denial of Service (DoS) attack can be executed by flooding malicious request into the vehicle's systems or communication networks. Some of the common Dos attacks in Autonomous vehicles are Sensor Request Overload and Network Flooding.

**Sensor Request Overload.** Autonomous vehicles rely on various sensors such as cameras, LiDAR, and radar to assess their environment surroundings. This attack is done by flooding these sensors with a large volume of data or signals or requests, causing them to become overloaded. This can lead to sensor malfunctions or inaccuracies, and these can tamper the vehicle's ability to make informed decisions.

**Network Flooding**. Autonomous vehicles depend on wireless communication networks for the exchange of data like receiving updates, sharing information, or accessing cloud services. Attackers can flood the network infrastructure with an excessive amount of requests which can overwhelm the network's capacity and causing delays or disruptions in critical communication channels.

*Man in the middle attack*

In a Man in the Middle attack, an attacker intercepts and manipulates communication between two parties without their knowledge. The attacker can eavesdrop on the communication and can be able modify data being exchanged or inject any malicious code which allows the attacker to gain unauthorized access to sensitive information or control over the communication channel. In the senario of autonomous vehicles this attack occurs when an attacker inserts themselves into the communication between the vehicle and other systems or devices, such as the vehicle's sensors, control systems, or network infrastructure. By intercepting the communication in the vehicle, the attacker can tamper the data being that is being transmitted to and from the sensors and manipulate the sensor readings or send malicious commands to the vehicle which can lead to dangerous consequences, such as the vehicle receiving false information, incorrect sensor inputs, or unauthorized control commands, which can compromise the safety and integrity of the autonomous vehicle system.

*Remote Code execution*

This is also known as RCE attack which enable attackers to execute any arbitrary code on a remote system. This means that an attacker can run their own code on a targeted system. An RCE can lead to unauthorized control over the system, compromising its security and allowing the attacker to perform malicious actions. In the scenario of autonomous vehicle impact of an RCE attack could be catastrophic as it has the potential to compromise the vehicle's functionality and compromise safety for both passenger and pedestrians. Using this attack can have the power to manipulate steering, acceleration, braking, or other critical functions while endangering the lives of passengers and others on the road. This attack can also enable unauthorized access to sensitive data and systems within the vehicle which includes personal information, location data, and communication systems.

**Attacks Based on Hardware Vulnerabilities**

Hardware vulnerabilities means weaknesses or flaws in the physical components of a system that can be exploited by attackers in order to gain unauthorized access, manipulate data and disrupt normal hardware operations. While These vulnerabilities can exist in various hardware elements such as processors, memory modules, input/output devices, firmware. As mentioned By (wierzynski & J, 2019), most of these vulnerabilities do not actually emerge from manufacturers they also lack some adequate updates with security.

*Sensor Spoofing*

Sensor spoofing is a hardware vulnerability that can have a serious impact on autonomous vehicles. Where sensor spoofing is all about messing with the vehicle's sensors which helps Autonomous vehicle to understand its surroundings. The aim of this attack is to trick the autonomous vehicle's perception system into making wrong decisions.So, When these sensors get spoofed, the vehicle gets confused about what's really happening around it. Then the attacker can make it think like there are obstacles where there are none or fail to detect real obstacles like other cars or pedestrians. By tampering with the sensor data, they mess up the vehicle's perception algorithms and make it navigate incorrectly or avoid collisions the wrong way. While the impact of the sensor spoofing attacks on autonomous vehicles is huge because it puts everyone from passengers to people on the street and other drivers at risk and can lead to accidents, mess up traffic flow, or even give unauthorized control over the vehicle to the attacker.

*Controller Area Network Bus Attack*

Controller area network is a communication system widely used in vehicles which enables differrent electronic components or micro controllers to communicate with each other. It acts as a central nervous system connecting various modules and components such as the engine control unit also known as ECU, brakes, dashboard and similar components . CAN helps in exchanging real time data and plays a very criritcal role in fucntioning of aunomous vehicles. As like any other hardawer or software CAN is also not immune to attack.

One of the common attack related to CAN is Contoller area network bus attack where this exploits the vulnerbailities in communication protocol to manipulate the flow of data in the bus. Attackers can attack CAN bus by using a 20 $hardware device which can snif the network to capture data packects and analyze CAN bus messages, reverse engineering the protocols used on the bus, and injecting malicious messages to manipulate the vehicle's systems. This can lead to various dangerous scenarios, such as altering sensor readings, controlling the vehicle's behavior, or even disabling essential systems like brakes or steering.

*ECU Firmware Manipulation*

The Term ECU has many different abbreviation Some manufacturers like to call Engine Control Unit and some of them call Electronic Computing unit. It is like a mini computer. Typically in an Autonomous vehicle there can be 600-2000 ECU's each of them have a different functionalities like Vision , Barrier Detection, Breaking, infotainment, mapping, detection of collision etc., They make their decision co-operatively with the central brain of the system in distributed manner. ECU firmware manipulation is an incredibly dangerous attack that poses a severe threat to the security of autonomous vehicles, By gaining unauthorized control over the vehicle's Electronic Control Units (ECUs), hackers or attackers can manipulate the software code within these vital components. This manipulation allows attackers to disrupt the vehicle's normal behavior and putting the driver's safety at risk. Moreover, it opens doors to privacy breaches, as sensitive information stored in the vehicle's systems becomes vulnerable. To address this issue, car manufacturers need to prioritize robust security measures. Implementing secure software updates, strong authentication mechanisms, and encryption can go a long way in protecting against ECU firmware manipulation.

### *GPS Spoofing*

GPS spoofing is a technique that can manipulate GPS signals received by autonomous vehicles. This attack involves broadcasting fake GPS signals that confuse the vehicle's navigation system into perceiving a different location than its actual position. It can be executed using specialized equipment which gives attackers the ability to manipulate the vehicle's actual location, speed, and direction. The consequesnces of GPS spoofing on autonomous vehicles can result in incorrect navigation, straying from planned routes and even collisions with obstacles.

To mitigate the risks associated with GPS spoofing. Advanced GPS receivers can be employed to identify and reject spoofed signals by analyzing their unique characteristics and comparing them with trusted sources. Furthermore by integrating additional sensors such as inertial navigation systems, cameras and lidar can provide cross-validation of the vehicle's position, ensuring consistency among different sensor inputs. To enhance the security of GPS signals secure communication protocols and encryption techniques can be employed to ensure the integrity and authenticity of the signals. By implementing these countermeasures, the vulnerabilities posed by GPS spoofing can be minimized (Joubert, G. Reid, & Noble, 2020).

## Security Architecture for Autonomous Vehicles

Security architecture for autonomous vehicles refers to the framework and measures that are put in place to ensure the security of the vehicle's systems and data which involves multiple layers of defense, including secure hardware components, good software security practices, secure communication protocols, and network infrastructure. The goal of Security architecture is to prevent Threats from compromising Vehicles operations . The security architecture also includes mechanisms for detecting and responding to security incidents and mitigating potential vulnerabilities.

### Components of Security Architecture

The security architecture for autonomous vehicles contains critical components Like Sensors, Communication channels and Software. The sensors that are used in autonomous vehicles which include radar, lidar, cameras and ultrasonic sensors, are vulnerable to cyberattacks such as spoofing or jamming. The Communication channels between autonomous vehicles and the surrounding infrastructure such as other vehicles and roadside units can be targeted by eavesdropping the data the is sent and recieved and denial-of-service attacks. The software responsible for decision-making within autonomous vehicles must be safeguarded against hacking and malware, while hardware components like computer systems, sensors, and actuators need protection against cyber intrusions (Chattopadhyay & Lam, 2018).

Apart form the above mentioned, there are some other measures to be implemented to ensure integrity and resilience. Secure hardware components like Trusted Platform Modules (TPMs) and Hardware Security Modules (HSMs) enhance overall system security. Secure coding practices and regular software updates mitigate vulnerabilities in software. Authentication and access control mechanisms ensure authorized access to critical functions and data. Secure communication protocols and encryption techniques protect data confidentiality and integrity during transmission. Intrusion detection and prevention systems monitor for suspicious activities in real-time. Data privacy measures safeguard sensitive user information. Over-the-Air (OTA) updates enable secure software and firmware updates. Fail-safe mechanisms and diverse technologies enhance system resilience. Continuous security monitoring and well-defined incident response plans detect and mitigate security breaches promptly.

## Vulnerbility Assessment and Penetration Testing

The future of autonomous vehicles depends on preventing attacks and minimizing cyber incidents to ensure the safety and security of passengers. One approach to identify vulnerabilities and enhance security is through penetration testing conducted by trained testers. Penetration testing, a part of autonomous vehicle security management, involves systematically assessing risks and minimizing security threats. Testers employ different testing approaches, including white-box, black-box, and grey-box testing, to optimize efficiency and realism while identifying vulnerabilities. Recommendations include the implementation of bug bounty programs, where individuals are incentivized to identify and report security flaws in exchange for rewards. This crowdsourced approach can help identify vulnerabilities that may have been missed during regular testing. Collaborating with managers is vital for documenting threats, designing security protocols, and creating policies. Furthermore, continuous monitoring, digital forensic analysis, and providing advice to improve systems are integral to maintaining a robust security architecture. Additionally, human support, remote assistance centers, and protecting the integrity of the information system are essential aspects of securing autonomous vehicles.

## Intrusion Detection and prevention Systems

**Detecting Techniques for intrusions**

Intrusion Detection and Prevention Systems (IDPS) are the systems that have been designed to detect and prevent unauthorized access and malicious activities in real-time. These systems play crucial role in the security architecture of autonomous vehicles.

As mentioned by (Basavaraj & Tayeb, 2022), there are various techniques that are employed by Intrusion Detection and Prevention Systems which are Signature based detection where this technique sinvovlve in comparing the network traffic against known attack signatures or patterns, So if the the signature get matched then the IDPS alerts and take action.Anomaly based and Heuristic based detections. They use the approach like identifying any deviation in the normal operations  and heuristic based detection relies on predefined algorithms in order to identify any suspicious behaviour.

**Real time monitoring approaches**

According to (Scarfone & Mell, 2007), there are few approaches that are commonly enouraged that are Network based Monitoring, Host based Monitoring, Behaviour based monitoring. As explained by (Scarfone & Mell, 2007), Network based Monitoring is based on the monitoring network traffic analysing and and identifying and suspicious activity while data transmission. Behaviour based monitoring observers behaviours of different processess and componets within a autonomous vehicle.Host based monitoring observes ans anlyzes the system logs, file itegrity and other system level indicators.

## Secure communication and Data Protection

**Vehicle to Vehicle Communication**

Vehicle-to-Vehicle (V2V) communication helps vehicles in exchanging data about traffic conditions and potential hazards. This communication protocol uses a Dedicated Short Range Communication (DSRC) and GPS receivers for effective data exchange . V2V communication also assists drivers in receiving real-time warnings about traffic and necessary actions that are needed to be taken. It uses radars and cameras to detect and avoid collisions. However, there are some challenges that include limited frequency band, communication protocol issues, cybersecurity threats, privacy concerns, and the need for human intervention. Despite these limitations, V2V communication has the potential to significantly improve road safety when properly implemented (Daddanala, Mannava, Tawlbeh, & Al-Ramahi, 2021).

**Vehicle and Infrastructure Communication**

V2I communication utilizes dedicated short-range communications (DSRC), cellular, and Wi-Fi technologies to enable vehicles to communicate with roadside infrastructure. DSRC operates in the 5.9 GHz band with a range of approximately 300 meters, providing high-bandwidth communication. V2I communication contains Safety applications that include collision and lane departure warnings, as well as red light alerts. It also contains Efficiency applications involve traffic signal coordination, public transportation priority, and parking guidance. DSRC offers reliable and efficient communication, while cellular and Wi-Fi provide wider availability but these have their limitations (Patrick Chan, P.E.).

<div align="center">**Regualation and Standards**</div>

**Regulations and Standards for Autonomous vehicles**

Regulations for autonomous vehicles are important because they help in ensuring safety, promote responsible development and deployment, protect consumer rights, and establish a framework for liability in case of accidents. In the same way Standards also help to harmonize practices, establishing common terminology and ensuring interoperability among autonomous vehicle systems and facilitating the innovation in industry.Some of Regulations and standards are ISO 26262: This functional safety standard focuses on road vehicles and provides guidelines for managing risks associated with electronic and electrical systems. It ensures that autonomous vehicles are designed, developed, and tested with safety in mind, addressing potential hazards and minimizing the risk of failures.

*SAE J3016*

This standard establishes common terminology for driving automation systems. Because It helps in classifying and understanding different levels of vehicle autonomy by enabling a clear communication and consistent understanding among the industry their respective regulatory bodies.

*NHTSA (FMVSS)*

FMVSS is also know as Federal Motor Vehicle Safety Standards which have been set by National Highway Traffic Safety Administration in the United States. FMVSS set the safety requirements for various aspects of motor vehicles, including autonomous vehicles while rating their crash worthiness, lighting, brakes and occupant protection.

*GDPR (General Data Protection Regulation)*

This is a regulation that is applicable in the European Union which mainly focuses on protecting the privacy and Personally Indentified Information know as PII data of individuals. It impacts autonomous vehicles Because they collect and process large amounts of data.This regulation helps in ensuring that appropriate safeguards are in place to handle and secure sensitive information.

*IEEE 2846*

This is a standard that focuses on the testing and validation of highly automated vehicles Because It provides a systematic approach to evaluate the performance, safety, and reliability of autonomous systems, ensuring that they meet the required standards before being deployed on public roads.

*California Autonomous Vehicle Regulations*

California state has also established their own regulations to govern the testing and deployment of autonomous vehicles within the state and These regulations include various aspects like licensing requirements, reporting obligations and safety standards to ensure safe operation on public roads.

**Standards for Cybersecurity in Autonomous Vehicles**

*ISO/SAE 21434*

According to U.S. Department of Transportation, 2022, This standard provides guidelines for cybersecurity in the whole automotive industry, which also include autonomous vehicles. It's main focus is on establishing a risk-based approach to identify and manage cyber threats, vulnerabilities, and incidents.

*ISO 26262*

According to Azianti, Liu, & Won , 2014, This standard specifically addresses the functional safety in the automotive industry which includes both regular automobiles and autonomous vehicles.

**Ethical and Legal Issues in Autonomous Vehicles**

**Ethical Considerations in Autonomous Driving Systems**

As very well said by (Lin, 2013) , "Sometimes good judgment can compel us to act illegally. Should a self-driving vehicle get to make that same decision? " Autonomous vehicles have some serious ethical implication such as Trolley problem which means If there is a situation of harming one person or a larger group then It is important to know how Autonomus cars should prioritize lives, Which could raise moral questions.Secondly it is Liability If there any event of accident which involves autonomous vehiles then determining a liability becomes a complicated issue. Now in this scenario who should be liable for the accident between A manufacturer, A software Developer, Owner of the vehicle to address this issue some clear frameworks has to be established for accountability.

Privacy and Data usage it is a known fact that autonmous vehicles collect and process large pool of data about their surroundings and passengers which raises the concern of privacy. It is crucial to make that the data is handled responsibly to protect the privacy of individuals and preventing unauthorised tracking or mointoring. Finally, Discrimination and fairness caould be a risk if these vehicles are programmed to discriminate certain race or social status. It is essential to develop regulation and safeguards which ensure unbiased treatment of all individuals preventing discrimination.

**Legal frameworks and Liability Issues**

Legal Frame works and liability issues play a very impotant role in deployment and development of Autonomous vehicles some of the legal issues are in an event of an accident, it is important to determine who is liable which is a huge challenge. According to (v & A, 2017), Article 8 paragraphs 1 and 5 of the Vienna Convention require that "[e]very moving vehicle or combination of vehicles shall have a [person] driver," and "[e]very driver shall at all times be able to control his vehicle." Article 13 paragraph 1 further requires that "[e]very driver of a vehicle shall in all circumstances have his vehicle under control so as to be able to exercise due and proper care and to be at all times in a position to perform all maneuvers required of him."

Contractual agreements must address the liability,intellectual property rights, data sharing and all other legal consideratiosn that are needed to be addressed among all the involved parties.

### Security Challenges in Autonomous Driving Systems

As autonomous vehicle use many technologies, hardware and software from the third parties It is important to adreess the isssue security as we all know every day newer security challenges arise as the technology is growing in fast pace.Some of the challages are Cyber attacks related to both hardware and software, Data integrity and privacy,Over-the-air updates and safety. These security challenges are needed to be addressed befor deployment of the autonomous vehicles in to the public infrastructure.

### Conclusion

**Summary**

In conclusion, As the rise of autonomous vehicles have some exciting possibilities for the future in the Automobile and transpotation industry there are some critical needs to address cybersecurity concerns such as vulnerabilities and potential threats in autonomous vehicles, such as unauthorized access, data breaches, denial-of-service attacks, and software vulnerabilities, pose significant risks to passenger safety and the integrity of the entire transportation infrastructure. In order To mitigate these risks, it is important to implement some good cybersecurity measures which include secure hardware and software, strong communication channels, intrusion detection and prevention systems and regular vulnerability assessments. There is also a need in Collaboration among industry stakeholders, researchers, and regulatory bodies is crucial in establishing comprehensive cybersecurity standards and best practices.  Apart the above mentioned there are also Ethical and legal considerations, such as the trolley problem and liability issues, must also be carefully addressed. By prioritizing cybersecurity in the development and before deployment of autonomous vehicles into the public infrastructure, we can ensure the safety, reliability, and public trust in this technology which can help revolutionize the transpotation industry.